\documentclass[aps,twocolumn,letterpaper,superscriptaddress,showpacs,preprintnumbers,amsmath,amssymb,prl]{revtex4}

\usepackage{ifpdf}
\usepackage{epsfig}
\usepackage{epstopdf}
\usepackage{graphicx}

\usepackage{graphicx}
\usepackage{dcolumn}
\usepackage{bm}
\usepackage{tikz}
\usetikzlibrary{matrix,arrows}
\usepackage{amsmath}

\begin{document}
\title{Multiphoton Ionization as a ``Clock'' to Reveal Molecular Dynamics with Intense Short X-ray Free Electron Laser Pulses
}

\author {L. Fang}
\email{lifang@slac.stanford.edu}
\author {T.~Osipov}
\author {B. Murphy}
\affiliation {Physics Department, Western Michigan University, Kalamazoo, MI 49008, USA}

\author {F. Tarantelli}
\affiliation {Dipartimento di Chimica, Universit\`{a} di Perugia, and ISTM-CNR, 06123 Perugia, Italy}

\author {E. Kukk}
\affiliation {Department of Physics and Astronomy,~University of Turku, 20014 Turku, Finland}

\author {J.P. Cryan}
\author {M. Glownia}
\affiliation {The PULSE Institute for Ultrafast Energy Science, SLAC National Accelerator Laboratory, Menlo Park, CA 94025, USA}

\author {P.H. Bucksbaum}
\affiliation {The PULSE Institute for Ultrafast Energy Science, SLAC National Accelerator Laboratory, Menlo Park, CA 94025, USA}
\affiliation {LCLS, Menlo Park, CA 94025, USA}

\author {R.N. Coffee}
\affiliation {LCLS, Menlo Park, CA 94025, USA}

\author {M. Chen}
\affiliation {Lawrence Livermore National laboratory, Livermore, CA 94550, USA}

\author {C. Buth}
\affiliation {Argonne National Laboratory, Argonne, Illinois 60439, USA}

\author {N. Berrah}
\affiliation {Physics Department, Western Michigan University, Kalamazoo, MI 49008, USA}

\date{\today}

\begin{abstract}

We investigate molecular dynamics of multiple ionization in N$_2$ through multiple core-level photoabsorption and subsequent Auger decay processes induced by intense, short X-ray free electron laser pulses. The timing dynamics of the photoabsorption and dissociation processes is mapped onto the kinetic energy of the fragments. Measurements of the latter allow us to map out the average internuclear separation for every molecular photoionization sequence step and obtain the average time interval between the photoabsorption events. Using multiphoton ionization as a tool of multiple-pulse pump-probe scheme, we demonstrate the modification of the ionization dynamics as we vary the x-ray laser pulse duration. 

\end{abstract}

\maketitle


Substantial efforts have been made both experimentally and theoretically to explore molecular dynamics during and after the interaction of the molecule with radiation, i.e., probe the electrons and nuclei behavior subsequent to absorption of photons~\cite{HHG1, HHG2, dynamics1, Frasinski_N2}. Among the number of factors determining this dynamics, the wavelength and intensity of the radiation are particularly important. Recent development of free electron lasers (FELs) has provided a novel source of short X-ray pulses with much higher intensity than available before~\cite{LCLS_recent1}. However, first experiments on molecular dynamics in this, hitherto unavailable range of parameters, have revealed that our understanding of the involved physics and chemistry is largely qualitative. We are lacking accurate models for quantitatively predicting the behavior of molecular systems that are subjected to intense X-rays interacting directly with atomic inner shells. The motivation of the present study is to remedy this situation by investigating molecular dynamics and fragmentation, subsequent to core ionization and core-hole relaxation, and by providing a  model and quantitative description of the interaction. These studies are of fundamental importance to molecular physics, physical chemistry and solid state physics. Furthermore, we expect that, with the rapid development of the FEL sources around the world, the need and the ability to accurately model the early phase of the molecular target breakup (that inevitably follows the intense X-ray pulse) becomes an essential part of many studies. 

The interaction of the intense X-ray FEL pulse with a molecule starts in the same way as for a lower intensity X-rays, i.e., with single-photon absorption and the ejection of a core electron. In
lighter elements such as nitrogen the resulting core hole is filled by Auger decay in a few femtoseconds (6.4 fs for N$_2$), with the ejection of a second electron (Auger electron). We will refer to this sequence as the photoionization-Auger (PA) cycle. Using conventional X-ray sources, the likelihood of the target molecule absorbing another photon before the dissociation is complete is negligible. In contrast,  the photon flux in the FEL pulse is so high that the dissociating molecule has a chance to absorb several X-ray photons in sequence, so that multiple PA cycles (a PAPA\ldots sequence) can take place in the molecular system within just one X-ray FEL pulse~\cite{DCH_Fang,  COLCLS_Berrah, N2LCLS_Hoener, N2LCLS_Cryan, Buth_N2}.

\begin{figure}[ht]
 \centerline{ \includegraphics[clip,width=1\linewidth]{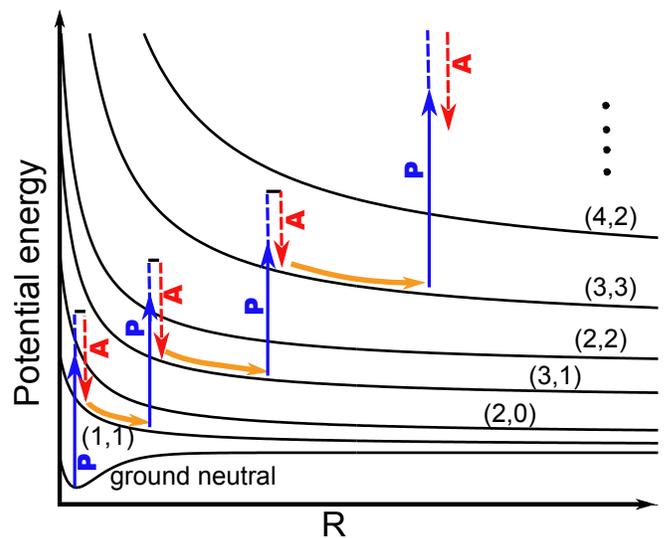}}
 	\caption{(Color online) Schematic of the PA sequences while molecules dissociate. The PA sequence shown here as an example is N$_2\rightarrow$(1,1)$\rightarrow$(3,1)$\rightarrow$(3,3). ``P'': photoionization; ``A'': Auger decay.}

  \label{fig_pot}
\end{figure}

In this work, we present an alternative to conventional pump-probe schemes~\cite{HHG2,HHG3,I2_Fang, dynamics2}, taking advantage of the femtosecond timescale of the multiple PA cycles in intense FEL pulses. With suitably chosen pulse intensity, the average interval between the PA cycles can be matched to the dissociation timescale, so that the observer follows the progress of the molecule along the repulsive potential energy surfaces (see Fig.~\ref{fig_pot}). We can regard the multiple PA cycles as an `‘intrinsic'' novel \textit{multiple-pulse pump-probe} mechanism where each subsequent cycle probes the development of the dynamics. Both the pumping and multiple probing come from a single X-ray pulse by sequential absorption. In the described multiphoton scheme, applicable to any ionization with intense X-ray-FEL pulses, the pump-probe delays are controlled by varying the FEL parameters.

In contrast to the pump-probe schemes based on two external pulses, the timing in the present scheme  is necessarily of statistical nature, since it is determined by intrinsic quantum events.
At each PA cycle, the kinetic energy (KE) of the emitted electrons reflects the molecular potential curves of the relevant charge states at a particular internuclear separation (Fig.~\ref{fig_pot}). The eventual kinetic energy release (KER) to the fragment ions, on the other hand, is determined by the entire dissociation dynamics. By combining these various observables with quantitative modeling, one can extract the averaged timing information and reconstruct molecular dynamics.
 
Thus, on one hand the PA processes provide a means for studying the electronic and structural
dynamics in the X-ray-molecule interaction. And on the other hand it is also an atomic intrinsic tool of quantum control of the interaction, since by manipulating the X-ray beam temporal profile one can modify the dynamics of sequential multiple ionization, opening or closing PA pathways to select desired outcomes~\cite{N2LCLS_Hoener, Buth_N2}.

In the present study, we examine the measured multiple ionization sequences of N$_2$ molecules with X-ray FEL pulses at equal pulse energies and various pulse durations, which result in different dynamics profiles of the interaction. We model the scenario of the interaction and quantitatively map the time dependence onto the average KE of the fragment ions. We extract the average time interval between photoabsorption events $\Delta t_{avg}$ as well as the corresponding internuclear separation ($R$) at which each step occurs in an ionization sequence. As expected, we observe that  $\Delta t_{avg}$ decreases with increasing intensity: a) at equal pulse energy which leads to a certain final charge state, $\Delta t_{avg}$ decreases with decreasing pulse duration, e.g., $\Delta t_{avg}$ decreases from 11-23 fs to 7-12 fs as the pulse duration decreases from 280 fs to 80 fs; b) at equal pulse duration, $\Delta t_{avg}$ decreases with increasing pulse energy, e.g., $\Delta t_{avg}$ decreases from 12-23 fs to 7-11 fs in X-ray spatial regions with increasing pulse energies where N$^{3+}$ to N$^{7+}$ are preferentially produced. This work takes advantage of sequential PA processes to \textit{reveal} the dynamics timing by developing a model, which is validated by the measurement of the fragment charge states. It shows the ability to quantify the correlation between ionization rates and KERs, and the possibility to resolve individual photoabsorption step in multiphoton ionizations. Also, this work presents  the modification of $\Delta t_{avg}$ with manipulation of the X-ray pulse duration and shows the variation of the ionization timing dynamics due to different pulse intensities within a single pulse.  

In the current experiment, the first PA step produces molecular dications N$_2^{2+}$ in either the ground state or excited states, with vacancies in outer molecular orbitals.  About 80\% of the electronic states of the dications that are formed in the Frank-Condon region, are dissociative~\cite{N2Auger_Agren, N2auger_Cryan}. Throughout this work, we will refer to the dissociation process N$_2^{n+m}\rightarrow$N$^{n+}$+N$^{m+}$ as the (n,m) channel. While the molecule breaks up, other PA cycles can take place locally at each of the atoms and thus the dissociation process may jump from the (1,1) to the (3,1) and then to the (3,3) channel. For a stable molecular dication the dissociation can be started in the (2,2) channel following the second PA process, then jumps to (4,2) and (4,4) channels by subsequent PA's. Similarly, higher charge states can be populated with further PA steps. The ionization sequences can be primarily defined by the final charge of the atomic fragments, but the sequences have common intermediate ionization steps. (See summary of possible pathways in Ref.~\cite{supporting}).

Starting with the ground state of neutral N$_2$, each photoionization step will lift the newly produced molecular ion to a potential curve of a higher charge state, as illustrated schematically in Fig.~\ref{fig_pot}. Each transition occurs at a different time and, once the dissociation starts, at a different $R$. For an ionization sequence,  $\Delta t_{avg}$ is determined by the inverse rate of the core ionization, which is given by the product of the core ionization cross-section and the X-ray flux. Hence, the higher the X-ray-FEL intensity, the shorter $\Delta t_{avg}$. This gives dissociation less time to proceed between ionizations and hence corresponds to a smaller $R$ for each PA step. The smaller the $R$ at which an ionization occurs, the stronger the influence of the neighboring atom on the one that hosts the photoionization process.

The experiment was performed in the high field physics (HFP) end station of atomic molecular and optical physics (AMO) hutch at the Linac Coherent Light Source (LCLS)~\cite{NeLCLS_Young, N2LCLS_Hoener, LCLS_Emma, LCLS_Bozek}. The photon energy used was 1100 ($\pm$15) eV, which is above the K-edge of neutral N and all charge states of N$_2$. The nominal pulse energy values from the gas detector monitors are measured upstream of the HFP chamber. In the interaction region the actual pulse energy is reduced by 65\%-85\% due to the beam transport loss~\cite{NeLCLS_Young, DCH_Fang, N2LCLS_Hoener}. The X-ray beam is focused by a pair of Kirkpatrick-Baez (KB) mirrors to an oval spatial profile with a major axis of 2.2 $\mu$m and a minor axis of 1.2  $\mu$m~\cite{DCH_Fang, N2LCLS_Hoener}. The pulse durations used were 280~fs, 110~fs, 80~fs and~7 fs, estimated by measuring the width of the electron bunch~\cite{Bane_measurepulsew, LCLS_Ding, LCLS_Dusterer}. The data shown in this work were recorded by an ion time-of-flight (iTOF) spectrometer, which is oriented perpendicularly to the plane defined by the polarization and propagation direction of the X-ray beam~\cite{N2LCLS_Hoener, N2LCLS_Cryan, LCLS_Bozek}.

To extract the KE of the ion fragments, we use a SIMION simulation to obtain a basis set of simulated iTOF spectral profiles corresponding to different charge states and KEs~\cite{Gessner_N2ion, supporting}. We adjust the weights of the basis for the best fit of the experimental spectra. These iTOF basis functions are generated in such a way that they automatically take into account all the non-trivial spectrometer transmission coefficients for different charge states and KE values. The average KEs obtained by this method give consistent results when using different basis sets and pools of initial values for fitting to the experimental spectra.

Knowing all the potential energy curves involved in a certain dissociation pathway with concurrent PA sequences, and $\Delta t_{avg}$, one can calculate the average KER in a channel (m,n). One also needs to know the time profile of the intensity of the FEL pulse; in this work we have approximated it by a square function ~\cite{LCLS_Dusterer} in which case the ionization rate is constant. The calculation can be performed also in reverse, obtaining KERs from experimentally determined KEs and then extracting R's and times for each preceding PA step. In this calculation we: i) assume the dominant contribution to the observed N$^{m+}$ ions is from the (m,m) channel, so that the KER is double that of the measured KE; ii) assume that there is one dominant pathway leading to ions of a certain charge; iii) assume, in the $R$ range of the dissociation of core-ionized molecules, core-hole potential curves have the same slopes as the potential curves of valence-hole states populated by the corresponding subsequent Auger decay process, i.e., assume that the Auger decay is instantaneous; iv) set $\Delta t_{avg}$ inversely proportional to the cross section of the atomic transitions; v) approximate potentials of high charge states with Coulomb potentials (a screening factor is applied to the (2,2) state). The validity of these assumptions and approximations is discussed in the supplementary material~\cite{supporting}.  

\begin{figure}[ht]
 \centerline{ \includegraphics[clip,width=1\linewidth]{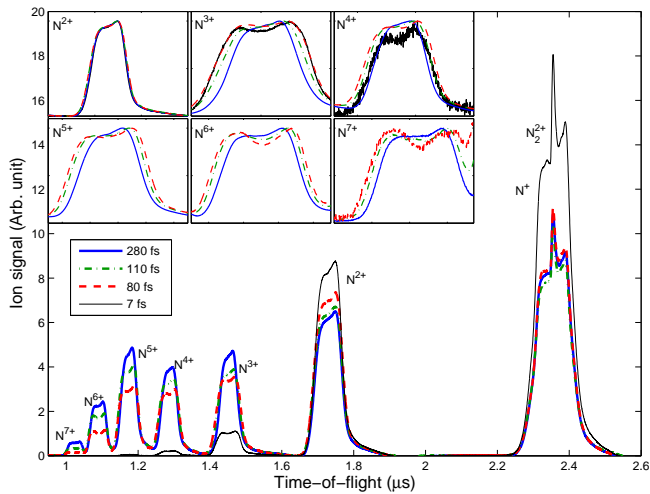}}
 	\caption{(Color on line) Ion spectra at different pulse durations. The pulse energy is 1.6--1.9 mJ (nominal values) for pulse durations $\geq$80 fs. The pulse energy for the 7 fs pulse is 0.2--0.3 mJ (nominal values). Main panel: spectra are normalized to the total ion signal of all charge states.  Insets: spectra for different charge states are normalized to the maximum of each spectral peak.}

  \label{fig_ion_highE}
\end{figure}
\begin{figure}[ht]
 \centerline{ \includegraphics[clip,width=1\linewidth]{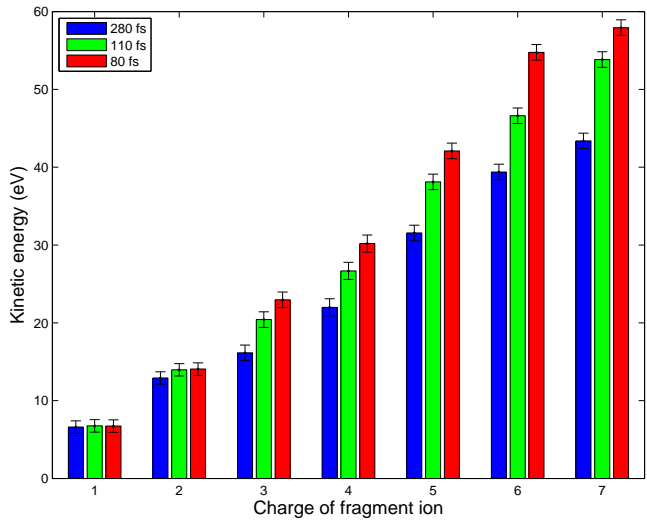}}
 	\caption{(Color on line) Average KE of ion fragments at different pulse durations, obtained from the spectra in Fig.~\ref{fig_ion_highE}.}

  \label{fig_KER}
\end{figure}

With intense X-ray pulses, we observe highly charged nitrogen fragment ions as a result of PA sequences, as shown in the main panel of Fig.~\ref{fig_ion_highE}. The spectra in Fig.~\ref{fig_ion_highE} are measured with different pulse durations at equal fluence, except for the 7 fs pulse duration data, for which the pulse energy is lower by 84\%. The 7 fs spectrum, from which the PA sequences are essentially missing and which is dominated by processes of molecular nature, is shown as a reference (a case where the least molecular dissociation is involved). As seen in the main panel of Fig.~\ref{fig_ion_highE}, higher X-ray flux results in higher charge states. This may suggest a molecular component to the $R$ dependent atomic core-electron absorption cross section which is changing during molecular dissociation. Also, the dynamics leading to the production of bare N$^{7+}$ appears to saturate. We observe that the pulse duration has a strong effect not only on the charge state branching ratios~\cite{N2LCLS_Hoener}, but also on the KE of each ion species. In the inset of Fig.~\ref{fig_ion_highE},  a wider structure corresponds to a larger average KE. For N$^+$ and N$^{2+}$, the peak widths are the same at the three different pulse durations, but starting from N$^{3+}$ the peak width decreases with increasing pulse duration. To quantify this trend, we obtain the average KE for different ion fragments at different pulse durations, which is shown in Fig.~\ref{fig_KER}. Since molecular ions or atomic fragments are detected after all ionization processes are over, the final KEs are the sum of contributions from all ionization steps involved. The KER shared by the ion fragments is determined by the $R$ where the PA or photoionization processes occur: the smaller the $R$, the larger the resultant KER. In other words, higher charges created at shorter separations convert more repulsive potential energy into the kinetic energy of the ions. As expected and seen in Fig.~\ref{fig_KER}, for N$^+$ and N$^{2+}$, the KEs are independent of the pulse duration, since these ions are mainly formed by PA processes in bound molecular states and the dissociation is not perturbed by further dissociation. For ionic fragments with charges $\geq$3+ the KE is larger for pulses with shorter duration but with similar fluence, i.e., a higher intensity (see data with pulse durations $\geq$80 fs). The observed increase of ion KE for shorter duration pulses supports the reasoning that higher photon intensity reduces $\Delta t_{avg}$. It is also seen from Fig.~\ref{fig_KER} that, for a given pulse duration the KE increases with the ion charge. Even at high charge states, e.g., for N$^{7+}$, we see an increase of KE compared with that of N$^{6+}$, reflecting the long range operation of Coulomb repulsion, and showing that the dissociation is not complete after N$^{5+}$ ionization.

\begin{figure}[ht]
 \centerline{ \includegraphics[clip,width=1\linewidth]{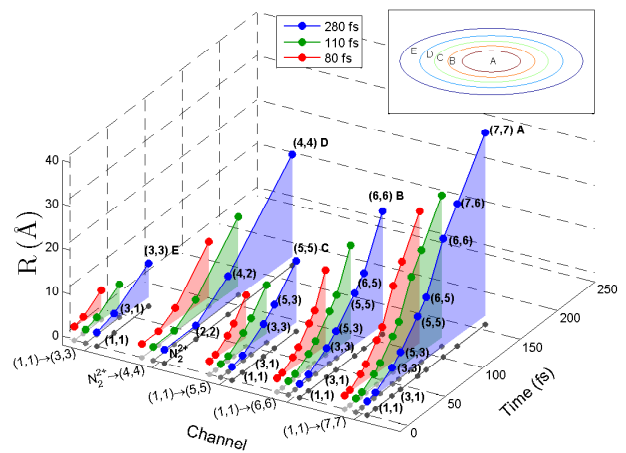}}
 	\caption{(Color on line) $R$ as a function of time for different PA pathways at different pulse durations. Each colored dot represents a photoabsorption event. The gray dots represent the projection of the event onto the channel-time plane. The channel axis labels indicate the states after the first PA sequence and the final states. The N$_2^{2+}$ potential curve used in the calculation is the $^1\Delta_g$ state from Ref.~\cite{N2curve_Wetmore}. Inset: schematic of the spatial intensity profile of the FEL beam; the labels correspond to the letter labels in the main panel, indicating that higher charge-state ions are preferentially produced in the region with higher pulse intensity.
} 

  \label{fig_tR_pw}
\end{figure}

As described previously, with the measured KE and using the above model, we extract $\Delta t_{avg}$ and the corresponding $R$ for various ionization sequences. This is done independently for different final charge states and the results are shown in Fig.~\ref{fig_tR_pw}. In this figure,  each block represents a pathway leading to the corresponding charge state; slices (or dots) of different colors within each block represent the ionization sequences for three different pulse durations. In our model, the time of the first ionization event is based on the average ionization rate, not the start of the FEL pulse, so the absolute starting times shown in Fig.~\ref{fig_tR_pw} are not equal. For a given final state, the ionization step occurs earlier with a shorter pulse than with a longer pulse. For instance, the $R$ (and $\Delta t_{avg}$) to reach N$^{3+}$ with 80 fs, 110 fs and 280 fs are 1.73$\pm$0.09 \AA (12 fs), 1.90$\pm$0.15 \AA (15 fs)  and 2.37$\pm$0.20 \AA (23 fs). We also find that, for a given pulse duration, the time to reach a specific intermediate charge state is different for ionization sequences leading to different final charge states. For example, using the KE of N$^{3+}$, N$^{5+}$, N$^{6+}$ and N$^{7+}$ at 280 fs, $R$ (and $\Delta t_{avg}$) for the transition (1,1)$\rightarrow$(3,1) are found to be 2.37$\pm$0.20 \AA (23 fs), 1.87$\pm$0.15 \AA (14 fs), 1.73$\pm$0.09 \AA (12 fs) and 1.69$\pm$0.08 \AA (11 fs), respectively. We thus see a decreasing $R$ and $\Delta t_{avg}$ for a given intermediate transition in ionization sequences associated with increasing final ion charges, as they are produced preferentially at increasing energies within a single pulse (see inset of Fig.~\ref{fig_tR_pw}).

The spatial intensity distribution in the focus leads to regions of high and low ionization rates and the measured KE is the convolution of these regions. Our ability to perform``pulse length resolved'' spectroscopy of dissociating N$_2$ molecules reveals that the involved molecular dynamics are influenced by both the spatial and temporal profile of the FEL pulse. Furthermore, the accuracy of the determined timing and $R$ for ionization steps depends on the potential energy curves used in the model. One could retrieve the information of the potential energy curves through measurements of the timing and $R$ which could be achieved in a standard pump-probe experiment. 

In summary, we investigate the ionization dynamics of N$_2$ molecules through sequential multiphoton core-level ionization and subsequent Auger decay.  Qualitatively, we expect and observe the molecular dynamics varying as a function of the X-ray intensity. With ion modeling, we are able to quantify the correlations between pulse durations, photoabsorption rates and fragment KEs. We obtain the average ion KE versus charge state and extract $\Delta t_{avg}$ and the corresponding $R$. With the multiphoton ionization scheme, we demonstrate the modification of the dynamics by varying the X-ray temporal profile. To completely disentangle the photoionization and the following Auger decay and further investigate the molecular dynamics, an in-depth modeling involving photoabsorption cross sections, Auger decay rates and molecular potential surfaces are necessary. Pump-probe experiments with X-ray pulses much shorter than the Auger decay time may in the future provide effective control of the involved dynamics, while also serving as a powerful FEL beam diagnostics tool. Finally, this work presents a quite general approach for treating ionization sequences accompanied by break-up of gas phase species, in cases of very high photon densities common at FEL sources.

\section{Acknowledgments}

We thank B. McFarland, C. Blaga, C. Bostedt, D. Rolles, E. P. Kanter, E. Hosler, J. D. Bozek,  L. DiMauro, M. G\"{u}ehr, M. Hoener, M. Messerschmidt, O. Gessner, O. Kornilov, S.R. Leone, S.T. Pratt and V. Petrovic for their earlier contribution. We thank C. L. Cocke and S.T. Pratt for their comments on the manuscript. This work was supported by the U.S. Department of Energy, Office of Science, Basic Energy Sciences, Division of Chemical Sciences, Geosciences, and Biosciences. E.K. acknowledges financial support from the Academy of Finland. C.B.~was supported by the Division of Chemical Sciences, Geosciences, and Biosciences of the Office of Basic Energy Sciences, Office of Science, U.S.~Department of Energy, under Contract No.~DE-AC02-06CH11357. The LCLS is funded by DOE-BES. We thank the LCLS staff for their assistance. 

\bibliographystyle{apsrev}
\bibliography{N2_ref}

\end{document}


\title{Supporting Materials
}



\begin{center}

\large

Supplemental Material to `Multiphoton ionization as a ``clock'' to reveal molecular dynamics with intense short X-ray free electron laser pulses'

\end{center}

\normalsize
\noindent\textbf{1. Pathways of ionization through photoionization and Auger decays processes}

The  photoionization and Auger decays (PA) pathways of N$_2$ are summarized in the chart below ($^*$ indicates states with 1s hole(s); the bold states and the double arrowed pathways are what we are interested in this work; horizontal lines correspond to ionization of one atom and vertical lines correspond to ionization of the other atom while the molecule dissociates; molecular dissociation starts at the second column): 
\begin{center}

\footnotesize{
{\begin{tikzpicture}[anchor=west]

\matrix (m) [matrix of math nodes, row sep=0.35em, column sep=0.00em, text height=0.05ex, text depth=0.05ex, nodes in empty cells]
{&\boldsymbol{(1,1)}\\
&\Downarrow\\
&(3,1)\Rightarrow&\boldsymbol{(3,3)}& & &\\
                          	  &\downarrow&\Downarrow & & &\\
			  &(5,1)\rightarrow&(5,3)\Rightarrow&\boldsymbol{(5,5)}  & &   \\
			  &\downarrow&\downarrow&\Downarrow & &\\
			  &(6^*,1)\rightarrow&(6^*,3)\rightarrow&(6^*,5)\Rightarrow&\boldsymbol{(6^*,6^*)} & \\
			  &\downarrow&\downarrow&\downarrow&\Downarrow& \\
			  &(7^*,1)\rightarrow&(7^*,3)\rightarrow&(7^*,5)\searrow& &\\
\rm{N}_2 & & & &(7^*,6^*)\Rightarrow&\boldsymbol{(7^*,7^*)} \\
			 &(7^*,2)\rightarrow&(7^*,4)\rightarrow&(7^*,5^*)\nearrow& &\\
			 &\uparrow&\uparrow&\uparrow&\uparrow&\\
			 &(6^*,2)\rightarrow&(6^*,4)\rightarrow&(6^*,5^*)\rightarrow&(6^*,6^*) &\\
			 &\uparrow&\uparrow&\uparrow& &\\
			  &(5^*,2)\rightarrow&(5^*,4)\rightarrow&(5^*,5^*) & &\\
                              &\uparrow&\uparrow& & &\\   
			 &(4,2)\Rightarrow&\boldsymbol{(4,4)}\\      
 			  &\Uparrow& & & &\\                       
                           \hspace{15pt}\rm{N}_2^{2+}\Rightarrow&\boldsymbol{(2,2)}& & & &\\
};

\path[->,double,font=\scriptsize,thick, bend left] (m-8-1) edge (m-1-2);
\path[->,double,font=\scriptsize,thick] (m-11-1) edge (m-17-1);

\end{tikzpicture}
}
}
\end{center}
\normalsize
\noindent Each PA step removes two electrons. The first PA step mainly doubly ionizes the neutral molecules in the ground state. The resultant dicationic ions are in the semi-bound N$_2^{2+}$ states or excited states which lead to dissociation. The latter process dominates following the first ionization by the X-ray pulses~\cite{N2LCLS_Hoener}.  The asymptotic limit of the dissociation after the first core ionization and Auger decay is either N$^{+}$+N$^{+}$ (the (1,1) channel) or N$^{2+}$+N (the (2,0) channel)~\cite{Stolte_N2}. Among the observed ions, the N$^+$ ions are mainly from the (1,1) dissociation~\cite{N2LCLS_Hoener}. Possible processes leading to the N$^{2+}$ are 1) symmetric dissociation of N$_2^{4+}$ following the PA process on ions in the bound states of N$_2^{2+}$; 2) symmetric dissociation of N$_2^{4+}$ following DCH Auger decay; and 3) (2,0) dissociation or PA process on N from the (2,0) dissociation. According to previous experiments~\cite{DCH_Fang, Benndorf_N2coincidence, Eberhardt_N2frag}, process 2) and 3) are minor contributions and we consider process 1) as the main cause of the N$^{2+}$ ions. Highly charged ions are produced by the PA process or photoionization alone on ions of lower charge states, as shown in the chart. For example, during the dissociation of (1,1) channel, the N$^+$ ion may absorb another photon and is further ionized by the subsequent Auger decay process, resulting in N$^{3+}$; this N$^{3+}$ ion can be ionized to higher charge states by further PA processes (see columns in the above chart). N$^{4+}$ without DCHs and N$^{5+}$ cannot ionize further by Auger decay process and thus double charge increase by one absorption step is not possible. However, sequential photoabsorption can still occur in these fragments as long as they retain electrons, in which case the ions N$^{5+}$, N$^{6+}$ and N$^{7+}$ will be produced. Similarly, the other atom in the molecule can be ionized to higher charge states (see rows in the above chart). 

The fast Auger decay mechanism plays an essential role here by refilling the core holes in few femtoseconds and allowing for the PAPA\ldots sequence to continue. With even higher pulse intensities, the core 1s orbitals can be depleted of electrons before the Auger decay occurs. Such double core hole (DCH) processes reveal novel physics~\cite{DCH_Fang, COLCLS_Berrah, N2LCLS_Cryan}, but they are not essential in investigating molecular dynamics and play a minor role in this study.

\vspace{10 pt}

\noindent\textbf{2. Extraction of average ion kinetic energy}

To extract the single ion kinetic energy (KE) value we carry out the following procedure. First, we model a set of ``basis'' time-of-flight (TOF) traces by using a well calibrated spectrometer geometry in SIMION for each observed charge state of nitrogen atom. Each of the TOF traces from the “basis set” corresponds to a small range of initial KE values of the given ion. Next, we take a linear combination of the basis TOF spectra that produces the best fit to the experimental TOF spectrum of the corresponding charge state. The resulting weights of this linear combination are used to obtain the ion distribution in the energy domain. This distribution is further integrated to get the average KE for a given charge state.  

\vspace{10 pt}
\noindent\textbf{3. Model for exacting the timing information of the ionization}

We apply several assumptions to the extraction of the timing information, which do not affect qualitatively the results of this work. First, as seen in the chart above, in the column of (m,n) with fixed n, the (n,n) state dominates the contribution to n+ ions because they require the fewest of photons and the contribution is double of that from the asymmetric states in the same column. Due to the latter, even though there is possibility that the (m,n) state is populated more than the (n,n) state (along the rows in the chart), the contribution from the (n,n) state could be still comparable with other states or even dominates. The above supports the assumption i) in the main text and the validity is confirmed with the observations that 1) there is no clear correlation between KEs of different charge states which is expected for pure (m,n) contributions where m$\neq$n; and 2) for the N$^+$ and the N$^{2+}$, the measured KEs agree with what is expected for dissociation on the (1,1) and (2,2) curve starting at the $R_e$ of N$_2$ ($\sim$1.1 \AA)~\cite{Coffee_N2}, which is consistent with the scenario of the first PA step. Based on the current available data, we cannot extract the kinetic energy distributions, from which the exact fragmentation channels may be retrieved. To fully resolve different pathways, covariance or coincidence experiments on molecular fragmentation are desired. Second, the assumption of time interval between photoionization steps (assumption iv) is validated by the X-ray temporal profile~\cite{LCLS_Dusterer}. Finally, there are pathways beside the ones used that lead to the observed charged ions. We have extracted the time and $R$ using different pathways involving extreme asymmetric states of (m,n) with m$\geq$(n-3) and find they lead to slightly smaller $R$ and time for each PA step compared with results using the assumed pathway. The assumptions and approximations applied in our model may add systematic errors to the absolute values of time and $R$, but remain a valid approach for showing the relative quantities in the dynamics.

In the model,  the average time between creation of N$^{i+}$ and the next photo-absorption is $\Delta t_{avg\_i}$ = (P*$\rho$*$\sigma_i$*$\omega_i$)$^{-1}$, where P is the photon flux, $\rho$  is the target density,  $\sigma_i$ is the atomic photoionization cross section, and $\omega_i$ is the weight of the particular transition. E.g. (1,1)$\rightarrow$(3,1) transition would have the weight of 2, because there are two N$^{+}$ ions that can be ionized to N$^{3+}$; while (3,1)$\rightarrow$(3,3) has weight of 1.  Given this $\Delta t_{avg\_i}$, the potential corresponding to the given molecular ion, and initial internuclear separation $R_i$  we find the increase of the kinetic energy $\Delta$E$_i$. The summation of all the $\Delta$E$_i$ along the complete multiple PA sequence gives the average kinetic energy release (KER) for a given path. The sum of KER over all ionization steps gives the average KER for a given charge state. The only unknown here is the product of the P*$\rho$ which is obtained by comparing to the experimental data. 

\vspace{10 pt}
\noindent\textbf{4. Multiple N$_2^{2+}$ states}

We only apply the potential curve of $^1\Delta_g$(1$\pi^{-2}$) for N$_2^{2+}$ to extract the ionization dynamics. According to FEL and synchrotron x-ray data, the dominant N$_2^{2+}$  states that dissociate populated by the PA sequence are $^1\Delta_g$(1$\pi^{-2}$) (24\%), $^1\Sigma_g$(1$\pi^{-2}$) (14\%) and $^1\Pi_g$(2$\sigma_u^{-1}$1$\pi_u^{-1}$) (9.8\%)~\cite{N2Auger_Agren, N2auger_Cryan} and the motion of ions on these different N$_2^{2+}$ states may vary. However, using these states, respectively, the results do not significantly deviate from that using the $^1\Delta_g$ state. Therefore, including these states won't change significantly the dynamics map and our results show the relative timing and position $R$ of the PA sequences.

\bibliographystyle{apsrev}
\bibliography{N2_ref}